\documentclass[prc,showpacs,nofootinbib]{revtex4}
\usepackage{feynmp,epsfig,graphics,color}

\def\tr{{\rm tr} \,}

\def\pslash{p \hspace{-1.7mm}/}
\def\barpslash{\bar{p} \hspace{-1.7mm}/}
\def\lslash{l \hspace{-1.7mm}/}
\def\uslash{u \hspace{-1.7mm}/}
\def\qslash{q \hspace{-1.7mm}/}
\def\barqslash{\bar{q} \hspace{-1.7mm}/}

\def\wslash{w \hspace{-1.7mm}/}

\def\w2{\tilde w^2}
\def\ws2{1}

\begin{document}
\title{Photoabsorption off nuclei with self consistent vertex corrections}
\author{F.\ Riek\footnote{Current affiliation: Cyclotron Institute and Physics Department, \\ Texas A\&M University, College Station, Texas 77843-3366, U.S.A.} and M.F.M. Lutz}
\affiliation{Gesellschaft f\"ur Schwerionenforschung (GSI),\\
Planck Str. 1, 64291 Darmstadt, Germany}
\author{C.L.\ Korpa,}
\affiliation{Department of Theoretical Physics, University of
Pecs, \\Ifjusag u.\ 6, 7624 Pecs, Hungary}
\date{\today}
\begin{abstract}
We study photoproduction off nuclei based on a self consistent and covariant
many body approach for the pion and isobar propagation in infinite nuclear matter.
For the first time the t-channel exchange of an in-medium pion is evaluated in the
presence of vertex correction effects consistently. In particular the interference pattern with
the s-channel in-medium nucleon and isobar exchange contribution is considered. Electromagnetic gauge
invariance is kept as a consequence of various Ward identities obeyed by the computation.
Adjusting the set of Migdal parameters to the data set we predict an attractive mass shift for the isobar of
about 50 MeV at nuclear saturation density.
\end{abstract}

\maketitle


\section{Introduction}

There is empirical evidence from photon nucleus absorption cross sections that the delta resonance changes
its properties in nuclear matter substantially already at
nuclear saturation density \cite{Hirata:Koch:Lenz:Monitz,photo-absorption,Richter}.
The microscopic description of the isobar self energy is a challenge
taken up by various groups \cite{Oset:Weise,Oset:Salcedo,Carrasco:Oset,Xia:Siemens:Soyeur,Korpa:Malfliet,Arve:Helgesson,Rapp,Korpa:Lutz:04,Post:Leupold:Mosel,Knoll,KoDi04,Hees:Rapp,Korpa:Lutz:Riek:2008}.
Naturally the study of the latter requires a solid understanding of the
pion spectral function in nuclear matter \cite{Ericson:Weise,Migdal,Nieves:Oset:Recio,Korpa:Lutz:Riek:2008} and references in \cite{Korpa:Lutz:Riek:2008}.

The phenomenological spreading potential \cite{Hirata:Koch:Lenz:Monitz} suggests a small repulsive
mass shift of the isobar together with an increase of its width. Also recent data on electroproduction of
isobars off helium three appear consistent with the latter interpretation \cite{Richter}.
With the exceptions \cite{Oset:Salcedo,Korpa:Lutz:04,KoDi04} model computations of the
isobar self energy claim results compatible with a small repulsive mass shift.
On naive grounds one may reject the works \cite{Oset:Salcedo,Korpa:Lutz:04,KoDi04}, that
predict a sizeable attractive mass shift for the isobar, as being unrealistic and incompatible
with nuclear photo absorption data. However, the situation is not clear cut. First, one may observe
that various detailed works \cite{Arve:Helgesson,Post:Leupold:Mosel} adjust their model parameters
as to reproduce the spreading potential \cite{Hirata:Koch:Lenz:Monitz} and therefore cannot be taken as
a microscopic confirmation of a repulsive isobar mass shift. Second, one should recall the
argument put forward in \cite{Oset:Salcedo,Carrasco:Oset,Korpa:Lutz:04} that the apparent mass
shift seen in photo absorption data is affected significantly by short range correlation
effects. Thus an attractive isobar mass shift cannot be ruled out, since the phenomenological
spreading potential \cite{Hirata:Koch:Lenz:Monitz} is effective in the sense that
the latter effects were not explicitly accounted for.

The purpose of this work is a study of the photoabsorption cross section off nuclei based on a self consistent and covariant many body approach for the pion and isobar propagation in infinite
nuclear matter that takes into account short range correlation effects consistently. We will apply
the  pion and isobar propagator as determined in \cite{Korpa:Lutz:Riek:2008} within a novel covariant approach
where vertex effects parameterized by Migdal's parameters are considered self consistently. Phenomenological soft
form factors  that would suppress vertex correction effects artificially are avoided in \cite{Korpa:Lutz:Riek:2008}.
Scalar and vector mean fields for the nucleon and isobar are incorporated consistently.

In the isobar region the t-channel pion-exchange process is known to define a sizeable background term for
the $\gamma \,p \to \pi^+ \,n$ reaction \cite{Ericson:Weise,Oset:Weise}. Thus it is crucial to consider the t-channel exchange of an
in-medium pion on equal footing as the in-medium exchange of the isobar when computing the photo absorption cross section of nuclei.
In this work, for the  first time, photo absorption is considered in the presence of short-range correlation effects in the $\gamma \,\pi \,\pi$, $\gamma \,N \,\Delta$, $\gamma\,\pi \,N\,\Delta $, $\pi\,N\,\Delta $ and $\pi \,N \,N$ vertices. Electromagnetic gauge
invariance is kept as a consequence of a series of Ward identities obeyed in the computation. In particular the interference
of the in-medium s-channel isobar exchange and the t-channel in-medium pion exchange is considered.

The set of Migdal parameters is adjusted as to obtain agreement with nuclear photo absorption data \cite{photo-absorption}. As a firm prediction we obtain an attractive mass shift for the isobar of about 50 MeV at nuclear saturation densities.
A comprehensive discussion of the relevance of various many-body effects is given.

\section{Photo absorption cross section}

We specify the isobar-hole model in its covariant form \cite{Nakano,Lutz:Migdal,Korpa:Lutz:Riek:2008}.
The interaction  of pions with nucleons and isobars is modeled by
the leading order vertices
\begin{eqnarray}
&& {\mathcal L}_{\rm int}= \frac{f_{N}}{m_\pi}\,\bar \psi
\,\gamma_5\,\gamma^\mu\,(\partial_\mu \vec \pi \,)\,\vec \tau
\,\psi+ \frac{f_{\Delta }}{m_\pi} \,\Big( \bar \psi^\mu
\, (\partial_\mu \vec \pi\,)\,\vec T\,\psi + {\rm h.c.} \Big)
\nonumber\\
&& \qquad +\,g_{11}'\, \frac{f^2_{N}}{m^2_\pi} \,
\Big(\bar  \psi \,\gamma_5\,\gamma_\mu\,\vec \tau \,\psi \Big)
\,\Big(\bar  \psi \,\gamma_5 \,\gamma^\mu\,\vec \tau \,\psi\Big)
\nonumber\\
&& \qquad +\, g_{22}'\, \frac{f^2_{\Delta }}{m^2_\pi} \, \Bigg( \Big(\bar
\psi_\mu \,\vec T \,\psi \Big) \,\Big(\bar \psi \,\vec T^\dagger
\psi^\mu\Big) + \left(\Big(\bar \psi_\mu \,\vec T \,\psi \Big)
\,\Big(\bar \psi^\mu \,\vec T \,\psi\Big)+{\rm h.c.}\right) \Bigg)
\nonumber\\
&&\qquad +\, g_{12}'\, \frac{f_{N}\,f_\Delta}{m^2_\pi} \,\Big(\bar \psi
\,\gamma_5\,\gamma_\mu\,\vec \tau \,\psi \Big) \left( \Big(\bar
\psi^\mu \, \vec T \,\psi\Big) +{\rm h.c.}\right)
\,,
\label{def-L}
\end{eqnarray}
where we use $T^\dagger_i\,T_j = \delta_{ij}- \tau_i\,\tau_j /3$ together with the free-space values  $f_{N}=0.988$ and
$f_{\Delta }= 1.85 $ in this work.  We consider
Migdal's short range correlation vertices  as introduced in
\cite{Nakano,Lutz:Migdal}, where it is understood that the local vertices are to be applied at
the Hartree level. The Fock contribution can be cast into the form
of a Hartree contribution by a simple Fierz transformation.
Therefore it only normalizes the coupling strength in
(\ref{def-L}) and can be omitted here.

\begin{figure}[b]
\begin{center}
\includegraphics[width=3cm]{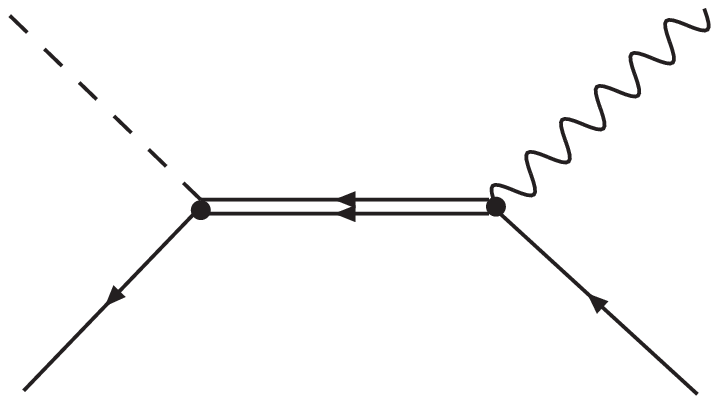} \includegraphics[width=3cm]{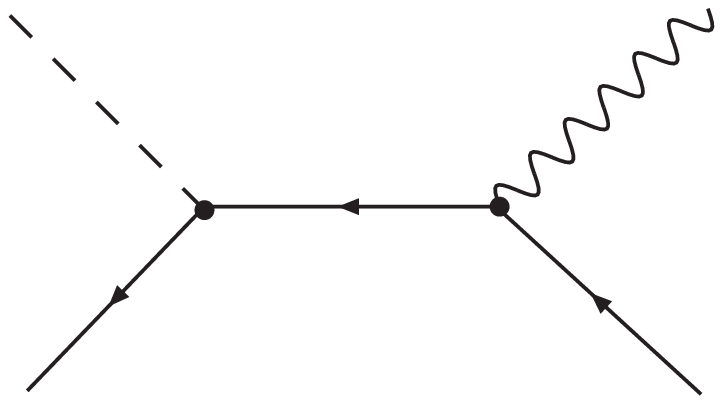}
\includegraphics[width=3cm]{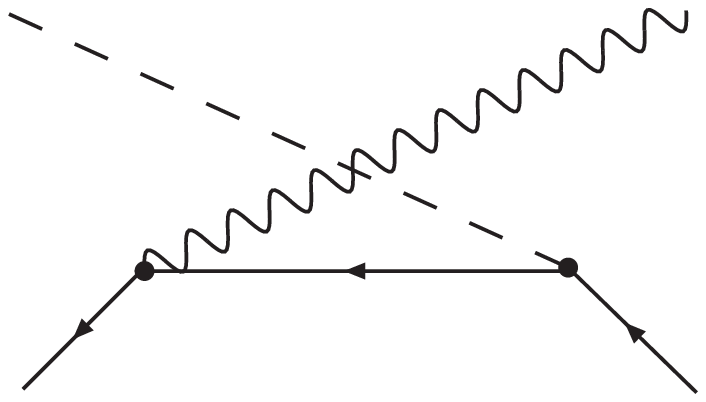} \includegraphics[width=1.8cm]{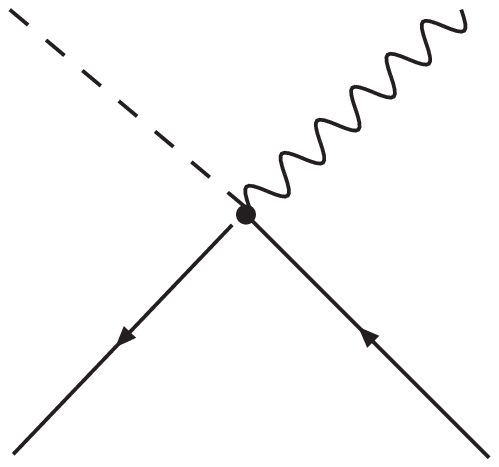}
\includegraphics[width=1.3cm]{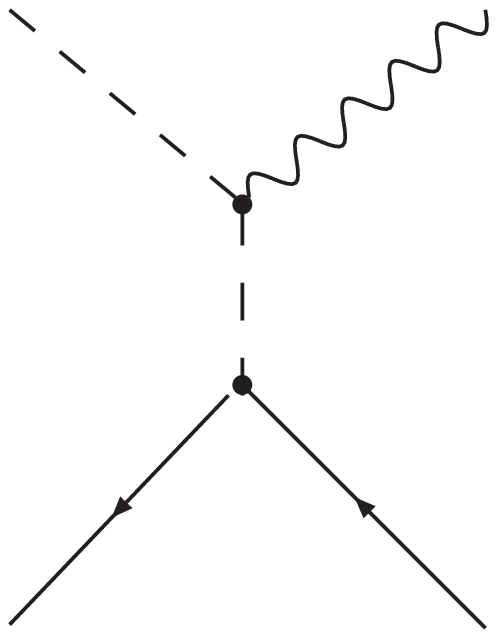}
\end{center}
\caption{Feynman diagrams for the photon induced pion production. }
\label{fig:1}
\end{figure}

\begin{figure}[t]
\begin{center}
\includegraphics[width=8cm]{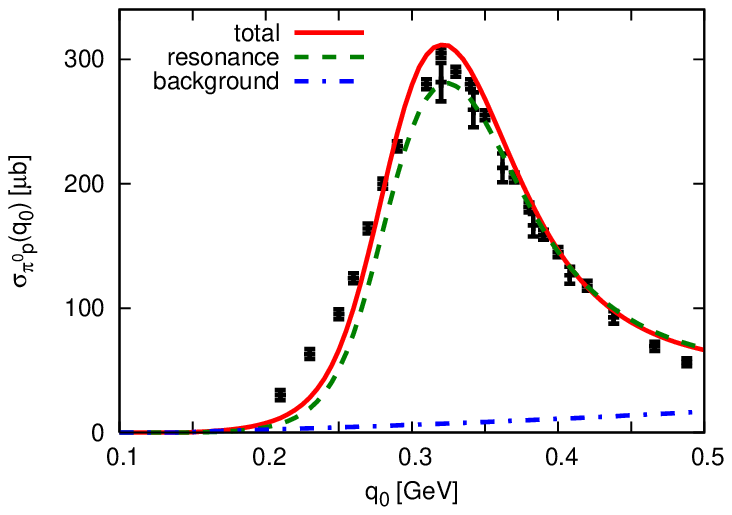} \includegraphics[width=8cm]{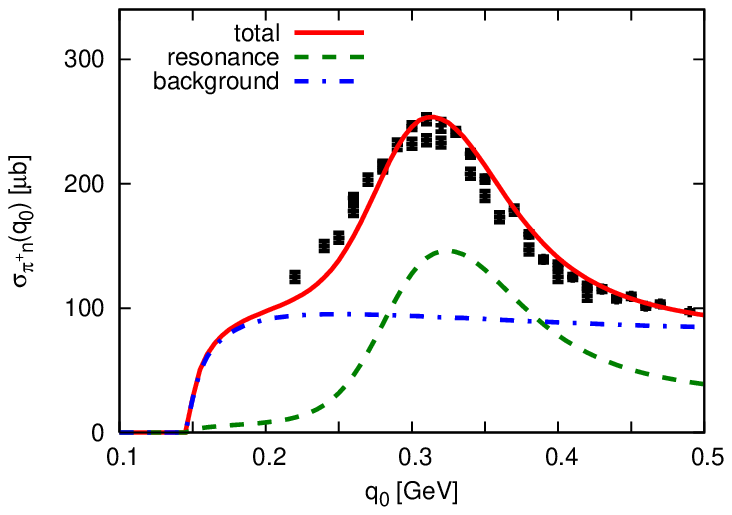}
\end{center}
\caption{The photo absorption cross section on the free proton evaluated in terms of
the isobar propagator of \cite{Korpa:Lutz:Riek:2008}. The data points are taken
from \cite{MacCormick:1996jz,vanPee:2007tw,Fujii:1976jg}. }
\label{fig:2}
\end{figure}

We supplement (\ref{def-L}) by leading order and relevant electromagnetic vertices
\begin{eqnarray}
&&{\mathcal L}_{\rm e.m.} =   - e\,A^\mu \, \bar \psi \,\frac{1+\tau_3}{2} \,\gamma_\mu \,\psi -e\,A^\mu\,(\vec \pi \times (\partial_\mu \vec \pi))_3
\nonumber\\
&& \qquad \;-\, e\,A_\mu \,\frac{f_N}{m_\pi} \,\bar \psi\,\gamma_5\,\gamma^\mu\,(\vec \tau \times \vec \pi)_3\,\psi
+\frac{i\,f_\gamma}{2\,m_\pi^2}\,\left( \epsilon_{\mu \nu \alpha \beta}\, F^{\alpha \beta}\,
  (\partial^\mu \bar \Psi^\nu)  \,T_3\,\psi + {\rm h.c.} \right)
\nonumber\\
&& \qquad \;+\, \frac{f'_\gamma}{m_\pi^2}\,\left(  F_{\mu \nu}\,
 (\partial^\mu \bar  \Psi^\nu)\,\gamma_5\,T_3\, \psi + {\rm h.c.} \right)\,,
\label{def-em-terms}
\end{eqnarray}
with the electromagnetic field strength tensor $F_{\mu \nu} = \partial_\mu A_\nu- \partial_\mu A_\nu $.
The magnetic and electric coupling constants $f_\gamma$ and $f_\gamma'$ are determined
from  the photon induced pion production cross section off the proton. We compute the cross sections as defined by the diagrams
of Figure \ref{fig:1}. The contribution of the u-channel isobar exchange is much suppressed and therefore neglected. The isobar propagator
is specified in \cite{Korpa:Lutz:Riek:2008}. It is modeled by a one-loop self energy describing the leading decay process of the isobar
into a pion and a nucleon. As illustrated in Figure \ref{fig:2} in the isobar region the photon-proton cross sections are reasonably well
described by the electric and magnetic coupling constants $f_\gamma = 0.012$ and $f'_\gamma = 0.024$.
Our values are close to the ones of Pascalutsa and Phillips, $f_\gamma \simeq 0.009$ and $f'_\gamma \simeq 0.021$  \cite{Pascalutsa:Phillips}.
While the neutral pion production is dominated by the s-channel isobar exchange contribution, the production of the charged
pion shows a sizeable background contribution. Following the arguments put forward in \cite{Korpa:Lutz:Riek:2008}, we consider the Lagrangian densities
(\ref{def-L}, \ref{def-em-terms}) to be effective and allow their parameters to enjoy a residual but smooth density dependence. The latter reflects
the dynamics of modes that are integrated out and therefore not treated explicitly here.

We compute the photo absorption cross section for an 'ideal' infinite nucleus.
Our studies will be based on the in-medium nucleon propagator parameterized in terms of scalar
and vector means fields:
\begin{eqnarray}
&& {\mathcal S}(p) = \frac{1}{\pslash-\Sigma_V^N\,\uslash -m_N + i\,\epsilon} + \Delta S(p)\,,
\qquad  m_N =m_N^{\rm vac}- \Sigma_N^S \,,
\nonumber\\
&& \Delta S (p) = 2\,\pi\,i\,\Theta \Big[p \cdot u-\Sigma_V^N \Big]\,
\delta\Big[(p-\Sigma^N_V\,u)^2-m_N^2\Big]\,
\nonumber\\
&& \qquad \qquad \times \,\Big( \pslash- \Sigma_V\,\uslash +m_N \Big)\,\Theta \Big[k_F^2+p^2-(u\cdot p)^2\,\big]\,,
\label{def-SN}
\end{eqnarray}
where the Fermi momentum $k_F$ specifies the nucleon density $\rho$ with
\begin{eqnarray}
\rho = -2\,\tr \,\gamma_0\,\int \frac{d^4p}{(2\pi)^4}\,i\,\Delta S(p)
= \frac{2\,k_F^3}{3\,\pi^2\,\sqrt{1-\vec u\,^2/c^2}}  \;.
\label{rho-u}
\end{eqnarray}
In the rest frame of the bulk with $u_\mu=(1,\vec 0\,)$ one recovers with (\ref{rho-u}) the
standard result $\rho = 2\,k_F^3/(3\,\pi^2)$. We assume isospin symmetric nuclear matter. The isobar propagator
$S_{\mu \nu}(w)$, is  the solution of Dyson's equation
\begin{eqnarray}
\!\!&&S^{\mu \nu}_0(w) =\frac{-1}{\wslash-m_\Delta+ i\,\epsilon} \left(
g^{\mu \nu}-\frac{\gamma^\mu\,\gamma^\nu }{3} -\frac{2\,w^\mu
\,w^\nu }{3\,m_\Delta^2} -\frac{\gamma^\mu \,w^\nu -w^\mu\,
\gamma^\nu}{3\,m_\Delta} \right)
\,,
\nonumber\\
&& S_{\mu \nu} (w) = S^{(0)}_{\mu \nu}(w-\Sigma^\Delta_V\,u)
+ S^{(0)}_{\mu \alpha}(w-\Sigma^\Delta_V\,u)\,\Sigma^{\alpha \beta}(w)\, S_{\beta \nu}(w)\,,
\label{Dyson}
\end{eqnarray}
where we allow for scalar and vector mean fields of the isobar with $m_\Delta = m^{\rm vac}_\Delta -\Sigma^\Delta_S$ as
developed in \cite{Lutz:Korpa:02,Lutz:Korpa:Moeller:2007,Korpa:Lutz:Riek:2008}. In \cite{Korpa:Lutz:Riek:2008} the pion and isobar self energy
$\Sigma^{\alpha \beta}(w,u)$ were determined in a self consistent and covariant many-body approach based on
the Lagrangian density (\ref{def-L}).
In this work we take the results of \cite{Korpa:Lutz:Riek:2008} and consider
the pion and isobar propagators as a function of $f_N, f_\Delta$ and $g_{ij}'$ and the mean field parameters for the nucleon and isobar.
For the details on the pion and isobar self energies we refer to \cite{Korpa:Lutz:Riek:2008}. It is the aim of the present work
to find sets of parameters that lead to a faithful representation of the nuclear photoabsorption data.

The computation of the total absorption  cross section is performed in the nuclear
matter rest frame. Fermi motion effects are considered. We express the cross section
\begin{eqnarray}
&&\sigma_{\gamma A}(q_0)= \frac{4}{\rho} \,\int_0^{k_F} \frac{d^3 p}{(2\pi)^3}
\, \frac{\Im \, A_{\gamma N  \to \gamma N}(q,p)}{2\,(p-u\,\Sigma_V)\cdot q} \,,
\nonumber\\
&& p_0=\sqrt{m_N^2+\vec p\,^2}+\Sigma_V\,, \qquad
q_0 = |\vec q \,|\,, \qquad  u_\mu=(1, \vec 0)\,,
\label{gam-sigma}
\end{eqnarray}
in terms of the imaginary part of the forward Compton amplitude $A_{\gamma N \to \gamma N}(p,q)$. We explore the role of
intermediate $\pi N$, $ Nh N$ and $\Delta h N$ states with
\begin{eqnarray}
&&\Im A_{\gamma \,N \to \gamma\, N} (q,p) =   \Im A^{(\pi)}_{\gamma \,N \to \gamma\, N} (q,p) +\Im A^{(ph)}_{\gamma \,N \to \gamma\, N} (q,p)
+\Im A^{(interference)}_{\gamma \,N \to \gamma\, N} (q,p)\,,
\nonumber\\
&& \Im A^{(\pi)}_{\gamma \,N \to \gamma\, N} (q,p) =\sum_{\lambda,i}\,{\rm Tr } \,
\int \frac{d^3 l}{(2\pi)^3}\,\frac{\Theta\big[ |\vec l \,|-k_F\big]}{16\,E_l}\,
\epsilon_\mu^\dagger(q, \lambda)\,T^{\dagger,i, \mu}_{\gamma \,N \to \pi\, N} (q,l ;w) \,
\nonumber\\
&& \qquad \qquad \qquad \quad\, \; \times \, \Big(\lslash + M \Big)\,
T^{i,\nu}_{\gamma \,N \to \pi \,N} (l,q;w)\,\epsilon_\nu(q, \lambda)\,
\Big(\pslash + M \Big)\,\rho^{(\pi )} (|w_0-l_0|,\, \vec w - \vec l\,) \Big|_{l_0 = E_l+\Sigma_V} \,,
\nonumber\\
&& \Im A_{\gamma \,N \to \gamma\, N}^{(ph)} (q,p) =  \sum_{\lambda,i}\,{\rm Tr } \,
\int \frac{d^3 l}{(2\pi)^3}\,\frac{\Theta\big[ |\vec l \,|-k_F\big]}{16\,E_l}\,
\epsilon_\mu^\dagger(q, \lambda)\,T^{\dagger,i,\alpha \mu}_{\gamma \,N \to ph\, N} (q,l ;w) \,
\nonumber\\
&& \qquad \qquad \qquad \quad\, \; \times \,  \Big(\lslash + M \Big)\,
T^{i,\beta\nu}_{\gamma \,N \to ph \,N} (l,q;w)\,\epsilon_\nu(q, \lambda)\,
\Big(\pslash + M \Big)\,\rho^{(ph)}_{\alpha\beta} (|w_0-l_0|,\, \vec w - \vec l\,)\, \Big|_{l_0 = E_l+\Sigma_V}\,,
\label{def-AgammapiN}
\end{eqnarray}
where the  '${\rm Tr}$' denotes the trace in Dirac and flavor space.  In this work we will neglect the interference term of the pion and
particle-hole contributions. The latter probes the product of the pion and particle-hole production amplitudes.
Furthermore, $M= m_N -\uslash \,\Sigma_V$ and $w= p+q$ and $E^2_l =m_N^2+\vec l\,^2$ with $E_l>0$.
We expect the most important contribution in (\ref{def-AgammapiN}) to result from the
intermediate $\pi N$ states, where we consider an effective
in-medium pion state characterized by its spectral distribution.
The effects from the nucleon-hole-nucleon ($Nh N$) and isobar-hole-nucleon ($\Delta h N$) states are described by a tensor spectral distribution.
This is possible since we consider only resonant contributions through the isobar s-channel process, for which the production amplitudes as implied
by (\ref{def-L}) are degenerate. Due to phase-space considerations the contribution from the $Nh N$ states is much larger as compared
to the one of the  $\Delta h N$ states, at least in the isobar region. This implies that this contribution will be roughly
proportional to $(g'_{12})^2$ and will become the more important the larger this value becomes.

We begin with a detailed exposition of the
pion and particle-hole spectral distributions $\rho_\pi(q)$ and $\rho_{ph}^{\alpha \beta} (q)$
required in (\ref{def-AgammapiN}).
The central building blocks, in terms of which they are expressed, are the short-range correlation bubbles
\begin{eqnarray}
&&\Pi_{\mu \nu}^{(Nh)}(q) = 2\,\frac{f^2_{N}}{m^2_\pi}
\,\int \frac{d^4 l}{(2\pi)^4}\,i\,\tr \,\Bigg( \Delta S(l)\,
\gamma_5 \,\gamma_\mu \,\frac{1}{\lslash +\qslash - M+i\,\epsilon}
 \,\gamma_5 \,\gamma_\nu \nonumber\\
&& \qquad \qquad \; +\, \frac{1}{2}\, \Delta S(l)\,
\gamma_5 \,\gamma_\mu \,\Delta S(l+q) \,\gamma_5 \,\gamma_\nu
\Bigg)  + (q_\mu \to -q_\mu ) \,,
\nonumber\\
&&  \Pi_{\mu \nu}^{(\Delta h)}(q) = \frac{4}{3}\,
\frac{f^2_{\Delta }}{m^2_\pi} \int \frac{d^4 l}{(2\pi)^4}\,i\,\tr
\,\Delta S(l)\,  S_{\mu \nu }(l+q) + (q_\mu
\to -q_\mu )  \,, \qquad
\Pi_{\mu\nu}(\bar{q})=\left(\begin{array}{cc}
\Pi_{\mu\nu}^{(Nh)}(\bar{q}) & 0 \\
0 & \Pi_{\mu\nu}^{(\Delta h)}(\bar{q})
\end{array}\right) \,, \qquad
\label{def-bubbles}
\end{eqnarray}
where '${\rm tr}$' denotes the trace in Dirac space. Note that the isobar-hole loop function in (\ref{def-bubbles}) is given in terms of
the in-medium isobar propagator as specified in (\ref{Dyson}) by the isobar self energy.
Details on the evaluation of the loop tensors (\ref{def-bubbles}) can be found in \cite{Korpa:Lutz:Riek:2008}.
For the spectral distributions $\rho_\pi(q)$ and $\rho_{ph}^{\alpha \beta} (q)$
of (\ref{def-AgammapiN}) we find
\begin{eqnarray}
&&\Pi_\pi(q) = -4\,\pi \left( 1+ \frac{m_\pi}{m_N}\right) b_{\rm eff} \,\rho
- \sum_{i,j=1}^2\,q_\mu\,\Big(
\Pi^{\mu \nu}_{ij}(q) + \Big[\Pi(q) \cdot \chi(q) \cdot \Pi(q)\Big]^{\mu \nu}_{ij}\Big) \,q_\nu \,,
\nonumber\\
&& g'_{\mu \nu}  =  \left(\begin{array}{cc}
g^{\prime}_{11} & g^{\prime}_{12} \\
g^{\prime}_{21} & g^{\prime}_{22}
\end{array}\right) g_{\mu \nu} \,, \qquad
 \qquad  \qquad \qquad \quad
 \chi^{\mu\nu}_{ij}(q)=\Big[ \Big(1-g' \cdot \Pi (q) \Big)^{-1} \!\!\cdot g' \Big]^{\mu\nu}_{ij} \,,
\nonumber\\
&& \rho_\pi(q) = -\Im \frac{1}{q^2-m_\pi^2-\Pi_\pi(q)}\,, \qquad \qquad \qquad \rho_{ph}^{\alpha \beta} (q) = \Im  \chi^{\alpha \beta}_{22} (q) \,,
\label{def-rhos}
\end{eqnarray}
where we recall the value $b_{\rm eff} \simeq -0.01$ fm from \cite{Korpa:Lutz:Riek:2008}. The latter value is needed to achieve
consistency with the low-density limit, where the pion self energy is determined by the s-wave pion-nucleon scattering length at $q_\mu =( m_\pi,0)$

\begin{figure}[t]
\begin{center}
\includegraphics[width=3cm]{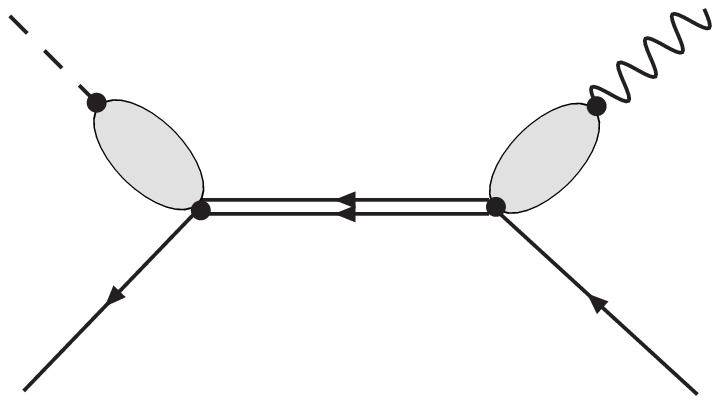} \includegraphics[width=3cm]{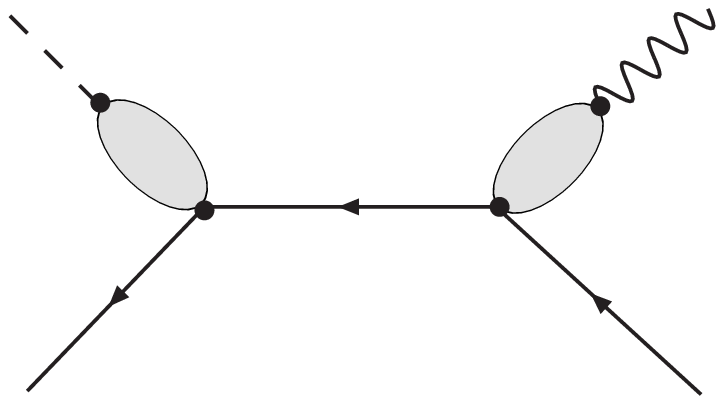}
\includegraphics[width=3cm]{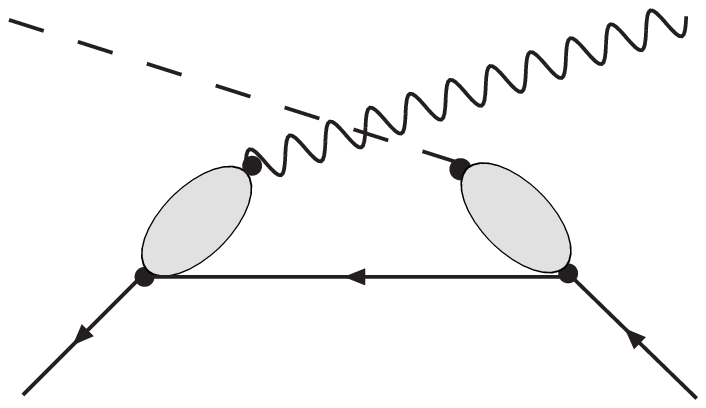} \includegraphics[width=1.8cm]{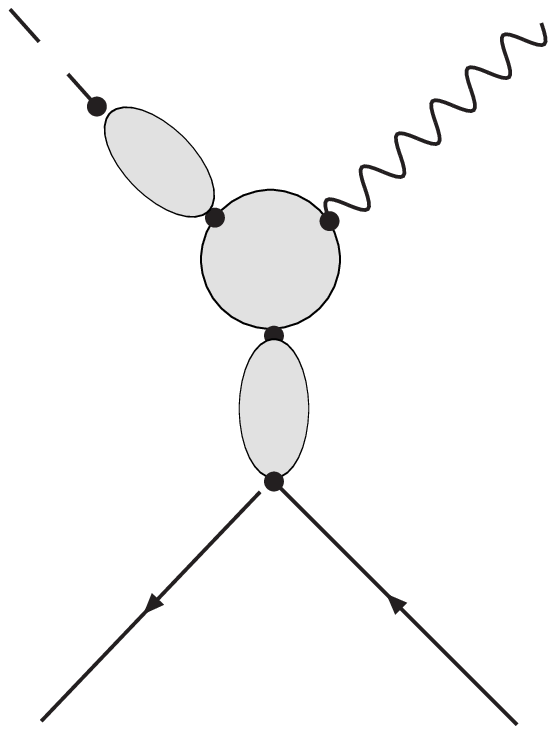}
\includegraphics[width=1.8cm]{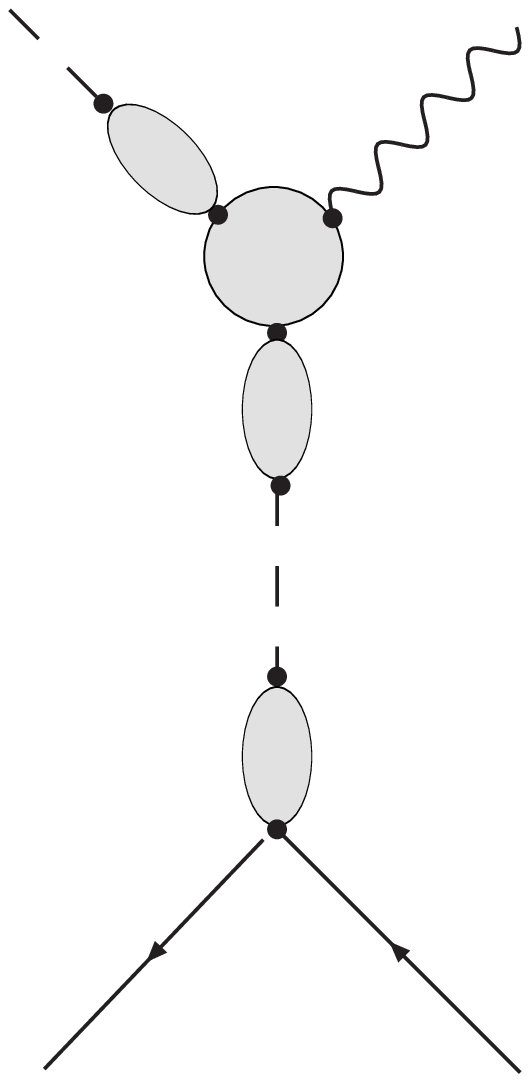}
\end{center}
\caption{Feynman diagrams that are considered for the in-medium $\gamma\,N \to \pi \,N$ process.}
\label{fig:3}
\end{figure}

We continue with the specification of the  pion and particle-hole production amplitudes $T^{i,\mu}_{\gamma \,N \to \pi \,N}$
and $T^{i,\alpha \mu}_{\gamma \,N \to ph \,N}$ in (\ref{def-AgammapiN}). In Figure \ref{fig:3} the in-medium generalization of the
diagrams of Figure \ref{fig:1} are shown, where short range correlation effects are considered in terms of various vertex functions.
A similar graphical representation holds for the particle-hole production amplitude, for which, however, we consider resonant contributions only.
More explicitly,  the in-medium $\gamma \,N \to \pi \,N $ and $\gamma \,N \to ph \,N $  amplitudes to be used in (\ref{def-AgammapiN}) take the form
\begin{eqnarray}
&&T^{\,\gamma \,N \to \pi\, N}_{i,\mu}(\bar{q},q;w)= \Lambda^{(\pi\, N\Delta)}_{i,\,\alpha}(\bar q )\,S^{\alpha\beta}(p+q)\,
\Lambda^{(\gamma N\Delta)}_{\mu\beta} (p,q)
\nonumber\\
&& \qquad \qquad \qquad \qquad +\,\Lambda^{(\pi\, NN)}_i(\bar q)\,S(p+q)\,\Lambda_{\mu}^{(\gamma \,NN)}(q)
 +\Lambda_{\mu}^{(\gamma \,NN)}(q)\,S(\bar p- q)\,\Lambda^{(\pi \,NN)}_i(\bar q)
\nonumber\\
&& \qquad \qquad \qquad \qquad +\,\Lambda^{(\gamma\,\pi \,NN)}_{i,\mu }(\bar q,q)
 +\Big[\Lambda^{(\pi\, NN)}_i(\bar q- q),\,{\textstyle{1\over 2}}\, \tau_3\Big]_- \, S_{\pi}(\bar q- q)\,\Lambda^{(\gamma\,\pi\,\pi)}_{\mu}(\bar q, \bar q-q)\,\,,
\nonumber\\
&& T_{i, \alpha \mu }^{\gamma \,N \to ph\, N} (\bar q, q;w)= \frac{f_{\Delta}}{m_{\pi}}\,T^\dagger_{i}\,g_{\nu \alpha}\,S^{\nu \beta}(p+q)
 \, \Lambda^{(\gamma N\Delta)}_{\mu \beta} (p,q) \,,
\label{def-TgammaN-piN}
\end{eqnarray}
where the various vertex functions  are subject to the Ward identities
\begin{eqnarray}
&& q^\mu\,\Lambda^{(\gamma\,N\,N)}_{\mu}(q) = e\,\frac{1+\tau_{3}}{2}\,\qslash\,,\qquad \qquad q^\mu\,\Lambda^{(\gamma\,N\,\Delta)}_{\mu \nu}(p,q) = 0\,,
\nonumber\\
&& (\bar q-q)^\mu\,\Lambda^{(\gamma\,\pi\,\pi)}_{\mu}(\bar q, q) = e\,\Big(\bar q^2- q^2- \Pi_\pi(\bar q)+ \Pi_\pi(q) \Big)\,,
\nonumber\\
&& q^\mu\,\Lambda^{(\gamma\,\pi\,N N)}_{i,\mu}(\bar q, q) = e\,\Big[{\textstyle{1\over 2}}\,\tau_3, \,\Lambda^{\pi NN}_i(\bar q)\Big]_-
-e\,\Big[{\textstyle{1\over 2}}\,\tau_3, \,\Lambda^{\pi NN}_i(\bar q-q)\Big]_-\,,
\label{def-Ward}
\end{eqnarray}
with the pion self energy $\Pi_\pi(q)$ of (\ref{def-bubbles}). Electro magnetic gauge invariance of the in-medium $\gamma \,N \to \pi\,N$
amplitude (\ref{def-TgammaN-piN}) is a consequence of  the identities (\ref{def-Ward}). The vertex  functions of (\ref{def-TgammaN-piN}) take the form
\begin{eqnarray}
&&\Lambda^{(\pi NN)}_i(\bar q) = \frac{f_{N}}{m_{\pi}}\,\tau_i \;\bar q_\mu \,\Big(g^{\mu \nu}
+\Big[\Pi (\bar q) \cdot \chi(\bar q)  \Big]_{11}^{\mu\nu} +\Big[ \Pi (\bar q)\cdot \chi(\bar q) \Big]_{12}^{\mu\nu}  \Big)\,\gamma_{5}\gamma_{\nu}\,,
\nonumber\\
&&\Lambda^{(\pi N\Delta)}_{i, \alpha}(\bar q) = \frac{f_{\Delta}}{m_{\pi}}\,T^\dagger_i \,\bar q_\mu \,\Big(g^{\mu \nu}
+\Big[ \Pi(\bar q)\cdot \chi (\bar q) \Big]_{22}^{\mu\nu} +\Big[ \Pi (\bar q)\cdot \chi(\bar q) \Big]_{21}^{\mu\nu} \Big)\,g_{\alpha \nu}\,,
\nonumber\\
&& \Lambda^{(\gamma\,N\,N)}_{\mu}(q)=e\,\frac{1+\tau_{3}}{2}\,\gamma_{\mu} \,,
\nonumber\\
&&  \Lambda^{(\gamma\,N\,\Delta)}_{\mu \alpha }(p,q)=
\frac{i\,f_\gamma}{m^2_\pi}\,T_3\,q^\tau\,\epsilon_{\mu \sigma \tau}^{\phantom{\mu \alpha \tau} \beta}\,
\Big( g_{\alpha }^\sigma\,p_\beta+ g_{\alpha \rho}\,
\chi^{\rho \kappa}_{22}(q)\,\Pi^{(\Delta h)}_{\kappa \sigma,\beta}(q)\Big)
\nonumber\\
&& \qquad \qquad \qquad -\,\frac{f'_\gamma}{m^2_\pi}\,T_3\,\gamma_5\, \Big( g_{\mu \alpha}\,(p \cdot q) -p_\mu\,q_\alpha \Big) \,,
\nonumber\\
&&\Lambda^{(\gamma\,\pi\,\pi)}_{\mu}(\bar q, q)=e\,\Big(q_{\mu}+\bar q_{\mu}
+\sum\limits_{ij=1}^{2} \Big( \Big[\Pi(q)+ \Pi(q) \cdot  \chi(q) \cdot \Pi (q)\Big]^{\nu \beta}_{ij}\,q_{\beta}
\nonumber\\
&&\qquad \qquad \qquad \qquad \qquad \qquad +\,\bar q_{\alpha}\, \Big[\Pi(\bar q)+\Pi(\bar q ) \cdot \chi(\bar q)\cdot \Pi(\bar q)\Big]^{\alpha\nu}_{ij} \Big)\,g_{\mu \nu}
\nonumber\\
&& \qquad \qquad \qquad -\,\sum\limits_{ijkl=1}^{2} \,\bar q_{\alpha}\,\Big[ 1+ \Pi(\bar q) \cdot \chi(\bar q)\Big]_{ik}^{\alpha \sigma}\,
\Pi^{(kl)}_{\mu,\sigma\tau}(\bar q,q)\, \Big[ 1+ \chi(q) \cdot \Pi(q)\Big]_{lj}^{\tau \beta}\,
q_{\beta} \Big)\,,
\nonumber\\
&&\Lambda^{(\gamma\,\pi\,NN )}_{i, \mu}(\bar q, q)=\frac{e\,f_N}{2\,m_\pi}\,[ \tau_3, \,\tau_i]_-\,\gamma_5\,\gamma_\nu\,\Big(
\sum\limits_{n=1}^{2} \, \Big[1+ \Pi (\bar q-q) \cdot \chi(\bar q- q)  \Big]^{\beta \nu}_{1n}\,g_{\mu \beta}
\nonumber\\
&&\qquad \qquad \qquad -\,\sum\limits_{jkl=1}^{2} \,\bar q_{\alpha}\,\Big[ 1+ \Pi(\bar q) \cdot \chi(\bar q)\Big]_{1k}^{\alpha \sigma}\,
\Pi^{(kl)}_{\mu,\sigma\tau}(\bar q,\bar q-q)\, \chi^{\tau \nu}_{lj}(\bar q-q)
\Big)\,,
\label{def-vertices}
\end{eqnarray}
with the loop tensors
\begin{eqnarray}
&& \Pi_{\mu \nu, \alpha }^{(\Delta h)}(q) = \frac{4}{3}\,
\frac{f^2_{\Delta }}{m^2_\pi} \int \frac{d^4 l}{(2\pi)^4}\,i\,\tr
\,\Delta S(l)\,  S_{\mu \nu }(l+q)\,(l+q)_\alpha+ (q_\mu \to -q_\mu )\,,
\nonumber\\
&&\Pi^{(11)}_{\mu, \alpha \beta}(\bar q,q)=2\,\frac{f^2_N}{m_\pi^2}\int\frac{d^4l}{(2\pi)^4}\,i\,{\rm tr} \,\Big\{
\gamma_{5}\,\gamma_{\beta}\,\Delta S(l)\,\gamma_{5}\,\gamma_{\alpha}\, \Big[
\Big( \frac{1}{\lslash+\barqslash-M+i\,\epsilon}
\nonumber\\
&& \qquad \quad  +\,\frac{1}{2}\,\Delta S(l+\bar q)\Big)\,\Gamma_\mu^{(\gamma \,N\,N)}(l+\bar q,l+q)\,
\Big( \frac{1}{\lslash+\qslash-M+i\,\epsilon}+ \frac{1}{2}\,\Delta S(l+q) \Big)
\nonumber\\
&& \qquad \quad  +\,\frac{3}{4}\,\Delta S(l+\bar q)\,\Gamma_\mu^{(\gamma \,N\,N)}(l+\bar q,l+q)\,\Delta S(l+q) \Big] \Big\}\,
  -(\bar q_\mu, q_\mu)\,\rightarrow\,-(\bar q_\mu,q_\mu) \,,
\nonumber\\
&&\Pi^{(22)}_{\mu, \alpha \beta}(\bar q,q)=\frac{4}{3}\,\frac{f^2_\Delta}{m_\pi^2}\int\frac{d^4l}{(2\pi)^4}\,i\,{\rm tr}\,\Big\{\Delta S(l)\,
S^{\kappa \sigma }(l+\bar q)\,g_{\alpha \kappa}
\nonumber\\
&& \qquad \quad  \times \,
\Gamma_{\mu,\sigma \tau}^{(\gamma\,\Delta \Delta )}(l +\bar q, l+q)\,S^{\tau \rho}(l+q)\,g_{\beta \rho} \Big\}
 -(\bar q_\mu, q_\mu)\,\rightarrow\,-(\bar q_\mu,q_\mu)\,,
\nonumber\\
&&\Pi^{(12)}_{\mu, \alpha \beta}(\bar q,q)=\frac{4}{3}\,\frac{f_\gamma}{e\,m_\pi^2}\,\frac{f_N\,f_\Delta}{m_\pi^2} \int\frac{d^4l}{(2\pi)^4}\,i\,{\rm tr}\,\Big\{
\gamma_5\,\gamma_\beta\,\Delta S(l)\,S^{\kappa\tau}(l+\bar q)\,g_{\alpha \kappa }
\nonumber\\
&& \qquad \quad  \times \,
\Gamma_{\mu,\tau}^{(\gamma\,N \Delta )}(l +\bar q, l+q)\,\Big( \frac{1}{\lslash+\qslash-M+i\,\epsilon}+ \Delta S(l+q) \Big) \Big\}
 -(\bar q_\mu, q_\mu)\,\rightarrow\,-(\bar q_\mu,q_\mu)\,.
\label{def-Pi-triangle}
\end{eqnarray}

Given the vertices (\ref{def-vertices}) the Ward identities (\ref{def-Ward}) follow if the loop tensors $\Pi_{ij}^{\mu, \alpha \beta}(\bar q,q)$  obey the reduced  Ward identities
\begin{eqnarray}
&&(\bar q-q)_\mu\,\Pi_{ij}^{\mu, \alpha \beta}(\bar q,q)= \delta_{ij}\, \Pi_{ii}^{\alpha \beta} (q) - \delta_{ij}\,\Pi_{ii}^{\alpha \beta} (\bar q)\,.
\label{def-reduced-Ward-identity}
\end{eqnarray}
The identities (\ref{def-reduced-Ward-identity}) hold provided that the $\gamma\,N\,N$, $\gamma\,N\,\Delta $ and $\gamma\,\Delta \, \Delta$ vertices in
(\ref{def-Pi-triangle}) satisfy the constraint equations
\begin{eqnarray}
&&(\bar p-p)^\mu\,\Gamma_\mu^{(\gamma \,N\,N)}(\bar p,p) = \barpslash -\pslash\,, \qquad  \qquad
(\bar p-p)^\mu\,\Gamma_\mu^{(\gamma \,N\,\Delta )}(\bar p,p) = 0 \,, \quad
\nonumber\\
&&(\bar p-p)^\mu \,\Gamma_{\mu,\alpha\beta}^{(\gamma\,\Delta \Delta )}(\bar p, p) = [S^{-1}]_{\alpha \beta}(\bar p)-[S^{-1}]_{\alpha \beta}(p) \,.
\label{def-rr-Ward}
\end{eqnarray}
We point out that the evaluation of $\Pi_{22}(\bar q,q)$ required the evaluation of
the diagrams of Figure \ref{fig:3}, where the photon couples to the intermediate pion-nucleon state building up the isobar self energy. This leads to a self consistency issue, since the latter requires the knowledge of the $\gamma \,\pi \,\pi$ vertex, which in  turn depends on $\Pi_{22}(\bar q,q)$.

To make progress we consider the following decomposition
\begin{eqnarray}
&& \Pi_{ij}^{\mu, \alpha \beta} (\bar q, q) = \frac{u_\mu  }{u\cdot (\bar q-q)}\,\delta_{ij}\,\Big(
\Pi_{ii}^{\alpha \beta} (q) -\Pi_{ii}^{\alpha \beta} (\bar q) \Big)
+\Delta \Pi_{ij}^{\mu, \alpha \beta} (\bar q, q)\,,
\nonumber\\
&& (\bar q-q)_\mu\,\Delta \Pi_{ij}^{\mu, \alpha \beta} (\bar q, q) =0 \,,
\label{Ward-triangle}
\end{eqnarray}
where we argue that the terms $\Delta \Pi_{ij}^{\mu, \alpha \beta} (\bar q, q)$ are suppressed by $1/m_N$ or $1/m_\Delta$
as compared to the first term in (\ref{Ward-triangle}). This is easily seen for the '$11$' term. The
$\gamma \,N\,N$ vertex takes the  form
\begin{eqnarray}
&&\Gamma^{(\gamma \,N\,N)}_\mu(\bar p,p) =\gamma_{\mu}+
\frac{2\,i\,f_\gamma}{e\,m^2_\pi}\,\gamma_5\,\gamma_{\nu}\,\epsilon^{\mu \tau \alpha \beta}\,\chi^{\nu \kappa}_{12}(\bar p-p)\,
\Pi^{(\Delta h)}_{\kappa \tau,\beta}(\bar p-p)\,(\bar p-p)_\alpha = \gamma_\mu \,,
\end{eqnarray}
where the vertex corrections vanish due to the anti symmetry of the $\epsilon $ tensor. A further possible contribution proportional to $f_\gamma'$ is
obsolete also. As a consequence  $\Delta \Pi_{11}^{\mu, \alpha \beta} (\bar q, q)$  enjoys a representation, which follows from
the one of $\Pi_{11}^{\mu, \alpha \beta} (\bar q, q)$ in (\ref{def-Pi-triangle}), upon the replacement
\begin{eqnarray}
\Gamma^{(\gamma \,N\,N)}_\mu=\gamma_\mu \to \gamma_\mu - u_\mu \,\frac{\barqslash-\qslash}{u\cdot (\bar q-q)}\,.
\end{eqnarray}
The $\gamma\,N\,N$ vertex in (\ref{def-Pi-triangle}) is sand witched between two nucleon propagators that are on-shell in the limit of a large nucleon mass. Since vector currents of massive particles are dominated by their zero component, our claim follows.
By analogy to the nucleon case, we expect the term $\Delta \Pi_{22}(\bar q,q)$ to be suppressed by $1/m_\Delta$ as compared to the first term in
(\ref{Ward-triangle}). Finally an explicit analysis of the term $\Delta \Pi_{12}(\bar q,q)$  reveals also its suppression by $1/m_N$. The
$\gamma\,N\,\Delta$ vertex in (\ref{def-Pi-triangle}) reads
\begin{eqnarray}
&&\Gamma^{(\gamma\,N\,\Delta)}_{\mu \alpha }(\bar p, p)=
i\,(\bar p-p)^\tau\,\epsilon_{\mu \sigma \tau}^{\phantom{\mu \alpha \tau} \beta}\,
\Big( g_{\alpha }^\sigma\,p_\beta+ g_{\alpha \rho}\,
\chi^{\rho \kappa}_{22}(\bar p-p)\,\Pi^{(\Delta h)}_{\kappa \sigma,\beta}(\bar p-p)\Big)
\nonumber\\
&&\qquad  \qquad \quad \;\;\; -\,\frac{f'_\gamma}{f_\gamma}\,\gamma_5\, \Big( g_{\mu \alpha}\,(p \cdot (\bar p-p)) -p_\mu\,(\bar p-p)_\alpha \Big)\,,
\label{}
\end{eqnarray}
where vertex corrections proportional to $f_\gamma'$ vanish identically.
The suppression of $\Delta \Pi_{12}(\bar q,q)$ follows upon an evaluation  of the appropriate trace in (\ref{def-Pi-triangle}).
Thus, in the following we neglect the terms  $\Delta \Pi_{ij}^{\mu, \alpha \beta} (\bar q, q)$ for $i,j=1,2$.
It is stressed that the term $u \cdot (\bar q-q)$  in  (\ref{Ward-triangle}) does not cause any kinematical  singularity for on-shell photons with
$(\bar q-q)^2 =0$.

\section{Numerical results and discussions}

\begin{figure}[t]
\includegraphics[scale=1.1]{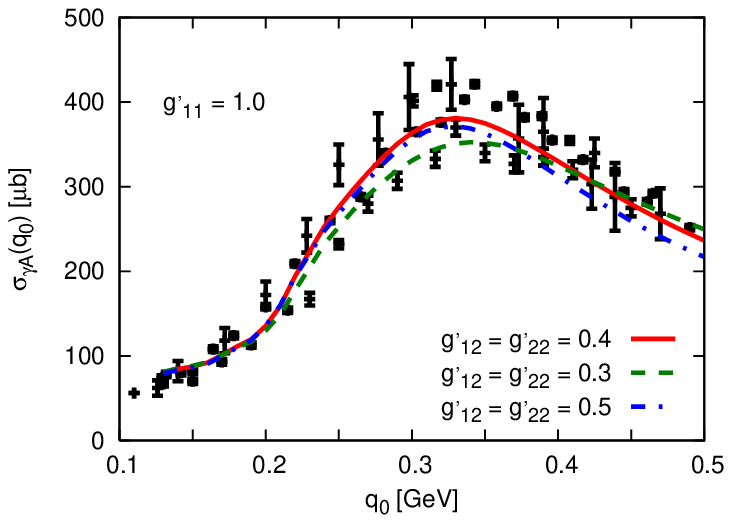}
\includegraphics[scale=1.1]{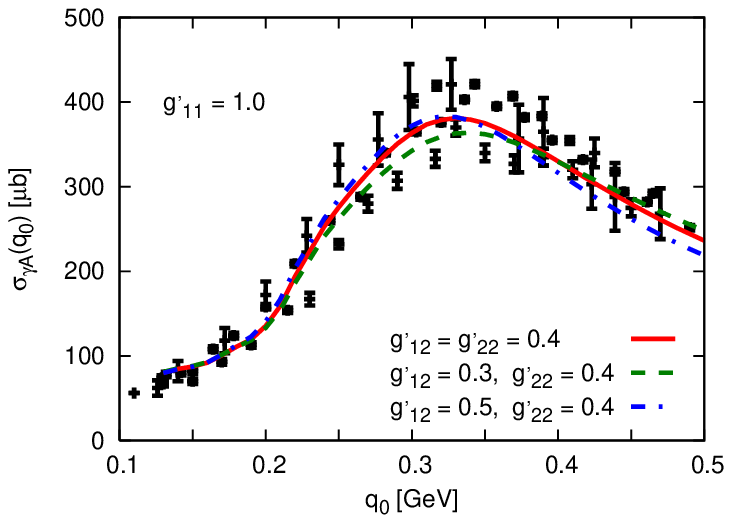}
\caption{Photoabsorption cross section for variations of $g^{\prime}_{12}$ and $g^{\prime}_{22}$.  We always keep $\Sigma^{\Delta}_{S}$ = -0.2 GeV and
$g^{\prime}_{11}=1.0$. The following values for $\Sigma_{V}^{\Delta}$ are used (corresponding to the legend from top to bottom): left figure: -0.09 GeV,  -0.10 GeV,  -0.07 GeV; right figure: -0.09 GeV, -0.10 GeV, -0.08 GeV. All parameters for 0.8 times saturation density. The data are taken from \cite{photo-absorption}.}
\label{fig:4}
\end{figure}

\begin{figure}[b]
\includegraphics[scale=1.1]{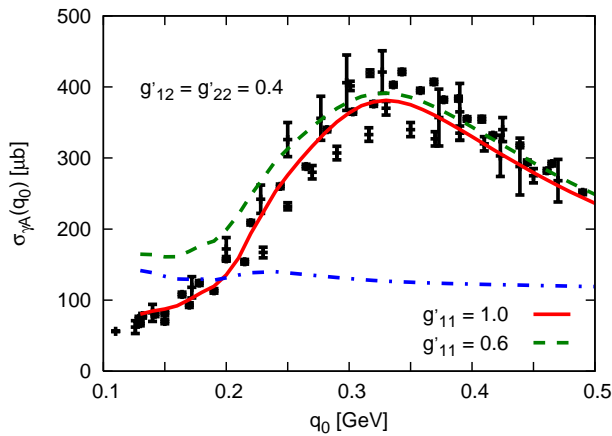}
\caption{Photoabsorption cross section for $g^{\prime}_{11}=1.0$ (solid line) and $g^{\prime}_{11}=0.6$ (dashed line).
We use $\Sigma^{\Delta}_{S}$ = -0.2 GeV and $\Sigma^{\Delta}_{V}$ = -0.09 GeV. In addition we show the background contribution for the run with $g^{\prime}_{11}$ = 0.6 (dash dotted line). The data are taken from \cite{photo-absorption}.}
\label{fig:5}
\end{figure}

We adjust the set of parameters to the photoabsorption data \cite{photo-absorption}. For the scalar and vector nucleon mean field we use the values
$\Sigma_S^N = 0.35$ GeV and $\Sigma_V^N = 0.29$ GeV at nuclear saturation density with $k_F =0.27$ GeV as
assumed also in \cite{Lutz:Korpa:Moeller:2007}. Following previous works \cite{Oset:Salcedo,Rapp} an
averaged density of 0.8 times saturation density is taken to compute the absorption cross section. The mean field parameters
for the nucleon are extrapolated down to that effective density by a linear ansatz.
We obtain a good description of the data \cite{photo-absorption} when using the following parameter set:
\begin{eqnarray}
&&\Sigma^{\Delta}_{S}=-0.25\,\rm{GeV}\,,\qquad\Sigma^{\Delta}_{V}=-0.11\,\rm{GeV}\,,\qquad
g'_{11}=1.0\,,\qquad g'_{12}=0.4\,,\qquad g'_{22}=0.4\,.
\label{standardparameter}
\end{eqnarray}
An extensive scan in the parameter space was performed. We assure that given our values for the nucleon mean fields there is a well defined
and localized region in parameter space that leads to an accurate reproduction of the photoabsorption data. A compilation of the results can be found in Figure \ref{fig:4} and Figure \ref{fig:5}. As it turns out we need a reduction of $f_\Delta $ and an increase of $f_\gamma$ as compared to their free-space values. Extrapolated linearly up to nuclear saturation density we derive a 15\% reduction of $f_{\Delta}$ and a 15 \% increase for $f_{\gamma}$.
Attempts to describe the data with no in-medium modifications of those parameters fails as the isobar turns too broad and
consequently the cross section too small.
A reproduction of the data set is possible also assuming a moderate reduction of $f_N$. However, this would require an even stronger medium modification
of the parameters $f_{\Delta}$ and $f_{\gamma}$. Changes in $f_{\gamma}^{\prime}$ have only a tiny influence on the
results so we keep this parameter at its free-space value.

In Figure \ref{fig:4} we study possible variations of the Migdal parameters $g'_{12}$ and $g'_{22}$ around the central
values $0.4$ of (\ref{standardparameter}). The magnitudes of the Migdal parameters are dependent to some extent on the subtleties
of the chosen approach. Thus we refrain from a detailed comparison with values obtained in different schemes. Keeping $g'_{11}=1.0$ and a scalar mean field for the isobar at $\Sigma^{\Delta}_{S}$ = -0.2 GeV, we readjust the magnitude for the vector isobar mean field. If we allow for variations larger that 0.1 in the Migdal parameters the cross section can no longer be reproduced accurately.  From Figure \ref{fig:4}
we see that with increasing values of $g^{\prime}_{12}$ and $g^{\prime}_{22}$ the shape of the cross section gets narrower.
The best description is obtained with a parameter set that delivers also the largest over all magnitude for the cross section.
Alltogether we arrive at the values of $g^{\prime}_{12}$ and $g^{\prime}_{22}$ to be round about 0.4.
In Figure \ref{fig:5} we illustrate the effect of lowering Migdal's parameter $g'_{11}$ down to 0.6. As seen in the figure
such a low value of $g_{11}'$ leads to a significant overshoot of the cross section at small photon energies. Though  the resonance
contribution itself is not affected much, the background contribution is enhanced strongly. This is shown by the dashed-dotted line which gives the result implied by all but the first diagram of Figure \ref{fig:3}.
We checked that variations of $g_{12}', g_{22}'$ or the isobar mean field parameters do not lead to a significant suppression of this contribution. The only mechanism to arrive at a smaller $g_{11}'$ would be
a significant reduction of $f_N$, however, at the price of an even larger reduction of $f_\Delta$.
Thus we arrive at a rather large value for $g_{11}'\simeq 1.0$.

\begin{figure}[t]
\includegraphics[scale=1.1]{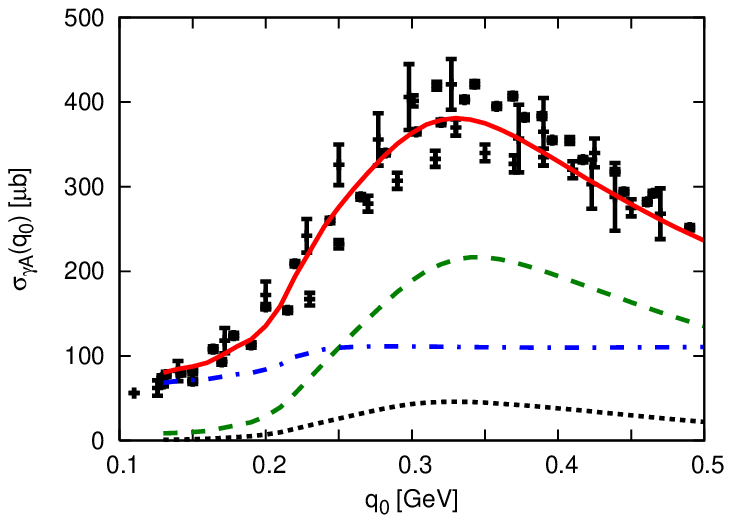}
\includegraphics[scale=1.1]{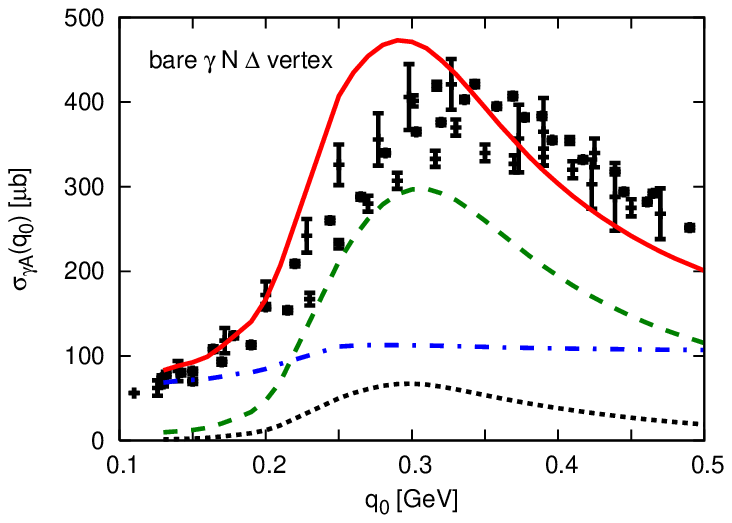}\\
\includegraphics[scale=1.1]{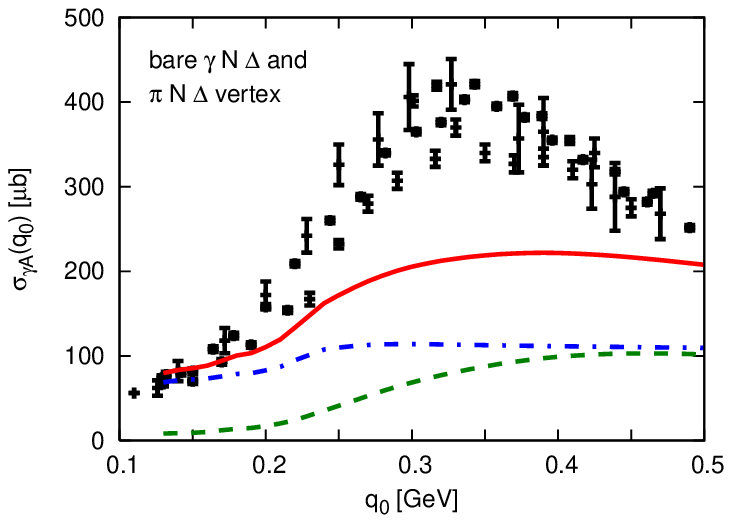}
\includegraphics[scale=1.1]{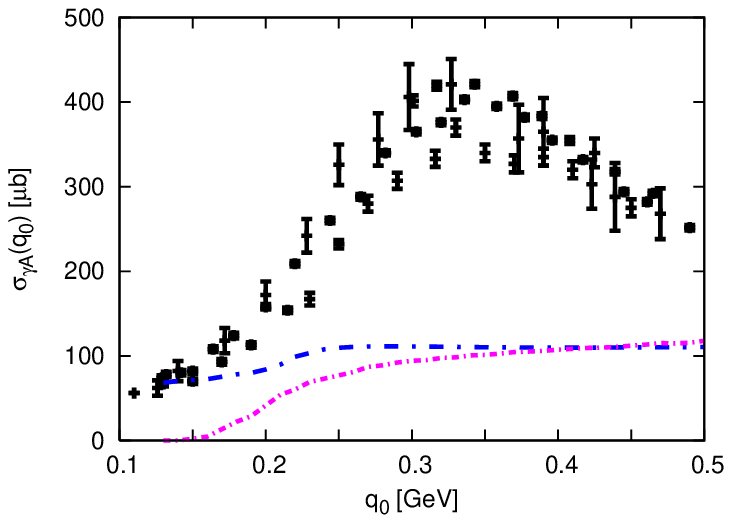}
\caption{Photoabsorption cross section using the  parameter set (\ref{standardparameter}). The upper-left panel shows the relevance of various contributions,
the upper-right panel the effect of short-range correlations in the $\gamma N\Delta $ vertex. The  lower-left panel
follows if bare   $\gamma N\Delta $ and $\pi N\Delta $ vertices are assumed.
The solid lines give the complete calculations, the dashed lines the resonance contributions, the  dashed-dotted lines the
background contributions and the dotted lines the two-particle hole contributions. The lower-right panel illustrates  the importance of in-medium effects on the background processes. The short-dashed dotted line provides the background contribution evaluated with bare vertices and a free-space pion. The data are taken from \cite{photo-absorption}. See the text for more details. }
\label{fig:6}
\end{figure}

In Figure  \ref{fig:6} we study the importance of various contributions and approximations. In the left upper panel the contributions of the resonance,
background and the one of the particle hole contributions are compared with the full result.
The background contribution, defined by  all but the first diagram of Figure \ref{fig:3}, is essentially flat and
delivers about 100 $\mu$b to the cross section. The resonance itself contributes about 200 $\mu$b in the peak while the
two-particle hole final states deliver an additional 50 $\mu$b. As can be seen when adding up all contributions
incoherently interference effects play a minor role only. We turn to the upper-right panel of Figure \ref{fig:6},
which illustrates the importance of vertex corrections.  The solid line of that panel gives our result implied by  the parameter set
(\ref{standardparameter}) but a  bare  $\gamma N \Delta$ vertex in the production amplitudes $T_{\gamma N \to \pi N}$ and
$T_{\gamma N \to ph N}$ of (\ref{def-AgammapiN}).  The neglect of short-range correlation effects in the $\gamma N \Delta $ vertex
implies a significant shift of the isobar strength about 50 MeV towards lower energies. Thus the apparent peak position seen in the
absorption cross section does not directly reflect the isobar contribution. A realistic prediction of the in-medium isobar mass
requires the proper consideration of such effects.
An even more dramatic influence of  short-range correlation effects is
documented by the lower-left panel of Figure \ref{fig:6}. Here we assume again the parameters set (\ref{standardparameter}) but also bare
$\gamma N \Delta $ and $\pi N \Delta $ vertices.  The pion and isobar propagators used are obtained within the self consistent and covariant
approach \cite{Korpa:Lutz:Riek:2008}, where correspondingly a bare $\pi N \Delta $ vertex was taken. This calculation
corresponds to the dashed lines in Figure 4 of \cite{Korpa:Lutz:Riek:2008}.
As anticipated by our previous study \cite{Korpa:Lutz:Riek:2008} a neglect of short-range correlation effects in the
$\pi N \Delta $ vertex leads to a much broader isobar which then translates into an almost flat photoabsorption cross section.
We finally turn to the lower-right panel of Figure \ref{fig:6}. Here we focus on the background contributions.
While the dashed-dotted line shows the full background contribution, the short dashed-dotted line gives the result for the background processes
implied when using a bare pion propagator and bare vertices in Figure 3. The vertex correction in the background terms are essential to keep our approach consistent. An approximative treatment in which the in-medium spectral distribution of the pion is neglected would lead to a strong underestimation of the
background processes. In this case the Pauli-blocking effect would cut away the low-energy cross section
as can be seen from Figure \ref{fig:6}. We emphasize that the consideration of such effects is crucial to arrive at a
realistic estimate for Migdal's parameter $g^{\prime}_{11}$.

It is interesting to compare our results with previous studies. We find a qualitative agreement with the results of \cite{Oset:Salcedo}, which claimed an attractive mass shift for the isobar in nuclear matter based on a perturbative and non-relativistic many-body approach.
This is in stark contrast to the more recent works \cite{Rapp,Hees:Rapp}, which claim small and repulsive mass shifts of the isobar in cold nuclear matter. The differences are traced to the neglect of important short-range correlation effects and the use of a soft and phenomenological form factor in the $\pi N \Delta $ vertex \cite{Rapp,Hees:Rapp}.

\section{Summary}

We presented a first computation of the nuclear photoabsorption cross section that considered the effect of short range-correlations effects in
the $\gamma \,\pi \,\pi$, $\gamma \,N \,\Delta$, $\gamma\,\pi \,N\,\Delta $, $\pi\,N\,\Delta $ and $\pi \,N \,N$ vertices. We applied
the self consistent and covariant many-body approach developed by the authors for
the $\pi N \Delta $ systems in the presence of short-range correlation effects. In particular
the in-medium interference of the s-channel isobar exchange and the t-channel pion exchange was evaluated consistently with an in-medium pion
propagator. It was shown that the latter plays an important role in the determination of Migdal's parameter $g'_{11} \simeq 1.0$,
for which we obtained a rather large value. An accurate reproduction of the photoabsorption data was achieved. Based on our analysis we predict an attractive mass shift of about 50 MeV  for the isobar in cold and saturated nuclear matter.

\vskip0.3cm

{\bfseries{Acknowledgments}}

\vskip0.3cm
F. R. acknowledges useful discussions with J. Knoll and would like to thank the FIAS (Frankfurt) for support.
C.L.K. would like to acknowledge financial support by the Hungarian Research Foundation
(OTKA grant 71989) and thank the G.S.I. (Darmstadt) and the K.V.I. (Groningen) for the kind hospitality.

\end{document}